\documentclass{moriond}

\newcommand{\eq}[1]{Eq.~\ref{eq:#1}}

\newcommand{\fig}[1]{Fig.~\ref{fig:#1}}

\newcommand{\mae}[3]{\langle#1|#2|#3\rangle}
\newcommand{\Mae}[3]{\bigl\langle#1\bigr|#2\bigr|#3\bigr\rangle}
\newcommand{\bra}[1]{\langle#1|}
\newcommand{\ket}[1]{|#1\rangle}

\newcommand{\df}{\mathrm{d}}
\newcommand{\img}{\mathrm{i}}

\newcommand{\tr}{\textrm{tr}}

\newcommand{\ga}{\gamma}

\newcommand{\eps}{\epsilon}
\newcommand{\ve}{\varepsilon}
\newcommand{\la}{\lambda}

\newcommand{\w}{\omega}
\newcommand{\balpha}{{\bar \alpha}}
\newcommand{\bbeta}{{\bar \beta}}

\newcommand{\cA}{{\mathcal A}}
\newcommand{\cB}{{\mathcal B}}
\newcommand{\cL}{{\mathcal L}}

\newcommand{\bn}{\bar{n}}
\newcommand{\bq}{{\bar{q}}}

\newcommand{\vC}{\vec{C}}
\newcommand{\vT}{\bar{T}}

\newcommand{\hS}{\widehat{S}}

\newcommand{\lp}{\tilde p}        
\newcommand{\ldel}{\tilde \delta} 

\newcommand{\C}{\mathrm{C}}      
\newcommand{\hard}{\mathrm{hard}}
\newcommand{\fin}{\mathrm{fin}}
\newcommand{\tree}{\mathrm{tree}}

\newcommand{\Photo}{}

\begin{document}

{\flushright \begin{tabular}{l}
DESY 16-087 \\
MIT--CTP 4805 \\
NIKHEF 2016-022\end{tabular}\\}

\vspace*{4cm}
\title{EMPLOYING HELICITY AMPLITUDES FOR RESUMMATION IN SCET}

\author{Ian Moult,$^1$ 
Iain W.~Stewart,$^1$
Frank J.~Tackmann,$^2$
Wouter J.~Waalewijn,$^{3,4}$
\footnote{Speaker} \vspace{1ex}}

\address{$^1$Center for Theoretical Physics, Massachusetts Institute of Technology, Cambridge, MA 02139, USA\\
$^2$Theory Group, Deutsches Elektronen-Synchrotron (DESY), D-22607 Hamburg, Germany\\
$^3$ITFA, University of Amsterdam, Science Park 904, 1018 XE, Amsterdam, The Netherlands\\
$^4$Nikhef, Theory Group, Science Park 105, 1098 XG, Amsterdam, The Netherlands}

\maketitle
\abstracts{
Helicity amplitudes are the fundamental ingredients of many QCD calculations for multi-leg processes. We describe how these can seamlessly be combined with resummation in Soft-Collinear Effective Theory (SCET), by constructing a helicity operator basis for which the Wilson coefficients are directly given in terms of color-ordered helicity amplitudes. This basis is crossing symmetric and has simple transformation properties under discrete symmetries.
}

\section{Introduction}

Precise predictions for Standard Model backgrounds are important to uncover new physics at the LHC. We focus on processes with hadronic jets, which receive large QCD corrections. There has been tremendous progress in calculating these corrections in fixed-order perturbation theory, using the spinor helicity formalism, color ordering techniques and unitarity based methods. Currently, NLO predictions are available for processes with a large number of jets and their computation has been largely automatized~\cite{NLO}.
Jet measurements often introduce a sensitivity to QCD effects at a scale $p$ well below  the partonic center-of-mass energy $Q$. Here $p$ corresponds to e.g.~the typical jet invariant mass or a veto on additional jets. The hierachy between $p$ and $Q$ leads to large  logarithms $\alpha_s^n\ln^m(p/Q)$ ($m \leq 2n$) in the cross section, that require resummation. 

\begin{figure}
 \hfill
\centerline{\includegraphics[width=0.7\textwidth]{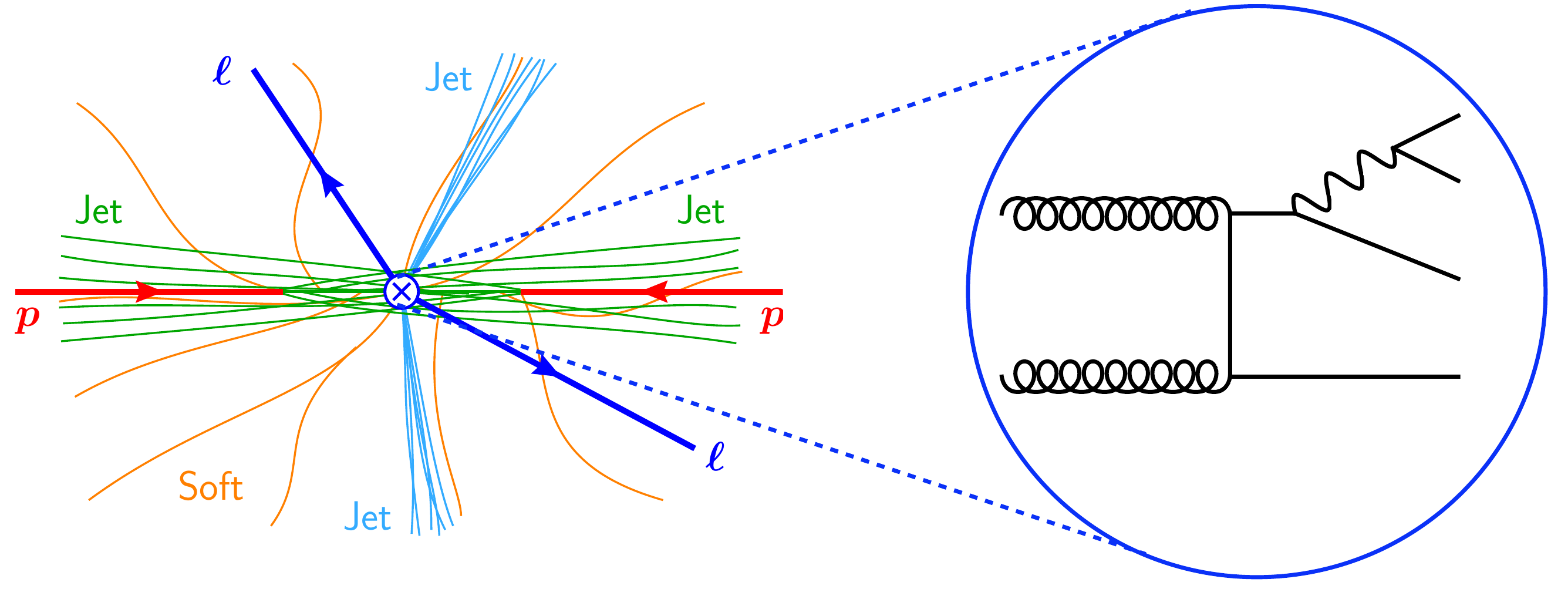}}
\hfill \vspace{-3ex}
\caption{Schematic LHC collision. The collinear (green and blue) and soft radiation (orange) are described dynamically in SCET. The hard scattering (zoomed in on the right) is encoded in matching coefficients.} 
\label{fig:hard}
\end{figure}

Soft-Collinear Effective Theory (SCET)~\cite{SCET} is an effective theory of QCD that enables resummation. It treats collinear and soft radiation (see \fig{hard}) as dynamical degrees of freedom,
\begin{equation}
 \cL_{\rm SCET} = \sum_n \cL_n + \cL_{\rm soft} + \cL_{\rm hard}
\,,\end{equation}
with $\cL_n$ the Lagrangian for collinear radiation in the light-like $n$ direction. The hard scattering is integrated out, due to the large virtuality of the momentum exchange, giving rise to $\cL_{\rm hard}= \sum_i C_i O_i$. Describing the spin content of operators $O_i$ with Dirac structures becomes cumbersome for complicated final states. We discuss a helicity operator basis which makes it easy to construct a complete basis and facilitates the matching from QCD onto SCET~\cite{Moult:2015aoa}.

In SCET resummation is achieved by decoupling the collinear and soft degrees of freedom in the Lagrangian~\cite{Bauer:2001yt}, leading to the following (schematic) factorized cross section 
\begin{equation} \label{eq:sigma}
\df\sigma =
\int\!\df \Phi(\{q_i\})\, M(\{q_i\})
\sum_{\kappa,\lambda}  \vC_{\la_1 \cdot\cdot(\cdot\cdot\la_n)}^\dagger (\{q_i\}) \hS_\kappa  \vC_{\la_1\cdot\cdot(\cdot\cdot\la_n)}(\{q_i\}) \otimes
\Bigl[ B_{\kappa_a} B_{\kappa_b} \prod_J J_{\kappa_J} \Bigr]
\,,\end{equation}
where the underlying Born process is $\kappa_a (q_a)\, \kappa_b(q_b) \to \kappa_1(q_1) \kappa_2(q_2) \cdots$.
$\df\Phi$ denotes the phase space integral and $M$ encodes the measurement on the hard kinematics. The restriction on collinear and soft radiation is encoded by the beam functions $B$, jet functions $J$ and soft function $S$. The matching coefficient $\vC_{\la_1\cdot\cdot(\cdot\cdot\la_n)}$ depends on the helicities $\la_i$ of the colliding partons and is a vector in color space. It cannot be combined with its conjugate, because the soft function sitting between them is a color matrix. As we will see in \eq{hard_match}, for our operator basis these Wilson coefficients are directly given in terms of color-ordered helicity amplitudes.

\section{Helicity operators}

We start by constructing quarks and gluon fields with definite helicity and then use this to construct our helicity operator basis. We will need the (conjugate) spinor with helicity $\pm$
\begin{equation} \label{eq:braket_def}
\ket{p\pm} = \frac{1 \pm \ga_5}{2}\, u(p)
\,, \qquad
\bra{p\pm} = \mathrm{sgn}(p^0)\, \bar{u}(p)\,\frac{1 \mp \ga_5}{2}
\,,\end{equation}
and the polarization vector for an (outgoing) gluon with momentum $p$ (with reference vector $k$)
\begin{equation}
 \ve_+^\mu(p,k) = \frac{\mae{p\!+\!}{\ga^\mu}{k+}}{\sqrt{2} \langle k\!-\!|p+\rangle}
\,,\quad
 \ve_-^\mu(p,k) = - \frac{\mae{p\!-\!}{\ga^\mu}{k-}}{\sqrt{2} \langle k\!+\!|p-\rangle}
\,.\end{equation}

The smallest building blocks of operators are the quark and gluon fields $\chi_{n,\w}$ and $\cB_{n,\w\perp}^\mu$, where $n = (1, \hat n)$ denotes the collinear direction and $\w = (1, -\hat n) \cdot p$ is the large component of its momentum $p$. These fields are invariant under collinear gauge transformations through the inclusion of Wilson lines. We define a gluon field of definite helicity by
\begin{equation} \label{eq:cBpm_def}
\cB^a_{i\pm} = -\ve_{\mp\mu}(n_i, \bn_i)\,\cB^{a\mu}_{n_i,\w_i\perp_i}
\,.\end{equation}
This definition is chosen such that that at tree level,
\begin{equation} \label{eq:B_tree}
 \Mae{g_\la^a(p)}{\cB_{i\la'}^{a'}}{0} = \delta_{\la,\la'}\, \delta^{a a'}\, \ldel(\lp_i - p)
 \,,\qquad
 \Mae{0}{\cB_{i\la'}^{a'}}{g_\la^a(p)} = (1 - \delta_{\la,\la'})\, \delta^{aa'}\, \ldel(\lp_i + p)
\,,\end{equation}
where the delta function $\ldel$ only fixes the large momentum component $\lp_i = \w_i n_i/2$. Exploiting that fermions come in pairs, we define fermion vectors currents of definite helicity
\begin{equation} \label{eq:jpm_def}
J_{ij+}^{\balpha\beta}
=  \frac{\sqrt{2}\, \ve_-^\mu(n_i, n_j)}{\sqrt{\phantom{2}\!\!\omega_i \,  \omega_j }}\, \frac{\bar{\chi}^\balpha_{i+}\, \gamma_\mu \chi^\beta_{j+}}{\langle n_i n_j\rangle}\,, 
\qquad
J_{ij-}^{\balpha\beta}
= -\, \frac{ \sqrt{2}\, \ve_+^\mu(n_i, n_j)}{\sqrt{\phantom{2}\!\! \omega_i \,  \omega_j }}\, \frac{\bar{\chi}^\balpha_{i-}\, \gamma_\mu \chi^\beta_{j-}}{[n_i n_j]}
\,,\end{equation}
which have similarly simple tree-level matrix elements.

It is now straightforward to write down the basis for a specific process. For example, 
for $gg q\bar q H$ the helicity basis consists of a total of six independent operators,
\begin{equation} \label{eq:ggqqH_basis}
O_{++(\pm)}^{ab\, \balpha\beta}
= \frac{1}{2}\, \cB_{1+}^a\, \cB_{2+}^b\, J_{34\pm}^{\balpha\beta}\, H_5
\,, \quad
O_{+-(\pm)}^{ab\, \balpha\beta}
= \cB_{1+}^a\, \cB_{2-}^b\, J_{34\pm}^{\balpha\beta}\, H_5
\,, \quad
O_{--(\pm)}^{ab\, \balpha\beta}
= \frac{1}{2} \cB_{1-}^a\, \cB_{2-}^b\, J_{34\pm}^{\balpha\beta}\, H_5
\,.\end{equation}
The symmetry factors in front of the operators account for identical fields. They ensure  the validity of \eq{Leff_me}, leading to a simple matching equation.

For specific processes, it is convenient to decompose the color structure of the Wilson coefficients using a color basis $T_k^{a_1\cdots\alpha_n}$, where $k$ runs over the allowed color structures. This yields
\begin{equation} \label{eq:Cpm_color}
C_{+\cdot\cdot(\cdot\cdot-)}^{a_1\cdots\alpha_n}
= \sum_k C_{+\cdot\cdot(\cdot\cdot-)}^k T_k^{a_1\cdots\alpha_n}
\equiv \vT^{ a_1\cdots\alpha_n} \vC_{+\cdot\cdot(\cdot\cdot-)}
\,.\end{equation}
For the $gg q\bar q H$ process a suitable color basis is given by
\begin{equation} \label{eq:ggqqH_color}
\vT^{ ab \alpha\bbeta}
= \Bigl(
   (T^a T^b)_{\alpha\bbeta}\,,\, (T^b T^a)_{\alpha\bbeta} \,,\, \tr[T^a T^b]\, \delta_{\alpha\bbeta}
   \Bigr)
\,.\end{equation}

\section{Matching}

For our helicity operator basis, the tree-level matrix element of $\cL_\hard$ is equal to the Wilson coefficient for the corresponding configuration of external particles,
\begin{equation} \label{eq:Leff_me}
\Mae{g_1g_2\cdots q_{n-1}\bar{q}_n}{\cL_\hard}{0}^{\tree}
= C_{+\cdot\cdot(\cdot\cdot-)}^{a_1 a_2\cdots\alpha_{n-1}\balpha_n}(\lp_1,\lp_2,\ldots,\lp_{n-1},\lp_n)
\,,\end{equation}
where $g_i \equiv g_{\lambda_i}^{a_i}(p_i)$ stands for a gluon with helicity $\lambda_i$, momentum $p_i$, color $a_i$, and analogously for (anti)quarks. This implies the tree-level matching equation
\begin{equation} \label{eq:matching_LO}
C_{+\cdot\cdot(\cdot\cdot-)}^{a_1\cdots\balpha_n}(\lp_1,\ldots,\lp_n)
= -\img \cA^\tree(g_1 \cdots \bar{q}_n)
\,,\end{equation}
where $\cA^\tree$ is the tree-level QCD helicity amplitude. 

In dimensional regularization, all loop corrections to the matrix element in \eq{Leff_me} are scaleless and vanish. These corrections consist of UV poles, which get renormalized, and IR poles, which cancel in the matching because SCET is an effective theory of QCD. This implies,
\begin{equation} \label{eq:hard_match}
C_{+\cdot\cdot(\cdot\cdot-)}^{a_1\cdots\balpha_n}(\lp_1,\ldots,\lp_n)
= -\img \cA_\fin(g_1\cdots \bar{q}_n) 
\equiv \frac{-\img \,
  \vT^{ a_1\cdots\bar\alpha_n}   \widehat Z_C^{-1} \vec {\cal A}_{\rm ren}(g_1\cdots \bar{q}_n)}{ Z_\xi^{n_q/2} Z_A^{n_g/2} }
\,.\end{equation}
Here $Z_\xi$, and $Z_A$ are the wave function renormalization of the quark and gluon field. $\widehat Z_C$ is the renormalization factor of the Wilson coefficient, which is a matrix in color space. At one-loop order $\cA_\fin$ is simply  the IR-finite part of the renormalized QCD helicity amplitude.

As an explicit example, we consider $gg q\bar q H$, for which the helicity operator basis was given in \eq{ggqqH_basis}. The color decomposition of the QCD helicity amplitudes into partial amplitudes is
\begin{equation} \label{eq:ggqqH_QCD}
\cA\bigl(g_1 g_2\, q_{3} \bq_{4} H_5 \bigr)
= \img\!\! \sum_{\sigma\in S_2} \bigl[T^{a_{\sigma(1)}} T^{a_{\sigma(2)}}\bigr]_{\alpha_3\balpha_4}
A(\sigma(1),\sigma(2); 3_q, 4_\bq; 5_H)
\!+\img\, \tr[T^{a_1} T^{a_2}]\,\delta_{\alpha_3\balpha_4} B(1,2; 3_q, 4_\bq; 5_H)
.\end{equation}
Using the color basis in \eq{ggqqH_color}, we can read off the Wilson coefficients. E.g.
\begin{equation} \label{eq:ggqqH_coeffs}
\vC_{+-(+)}(\lp_1,\lp_2;\lp_3,\lp_4;\lp_5) =
\left(\!\!\!\begin{tabular}{c}
   $A_\fin(1^+,2^-;3_q^+,4_\bq^-; 5_H)$ \\
   $A_\fin(2^-,1^+;3_q^+,4_\bq^-; 5_H)$ \\
   $B_\fin(1^+,2^-;3_q^+,4_\bq^-; 5_H)$ \\
\end{tabular}\!\!\!\right)
\,.\end{equation}
Charge conjugation invariance halves the number of independent Wilson coefficients.

\section{Properties}

Our operator basis is automatically crossing symmetric, because the gluon fields $\cB_{i\pm}$ can absorb or emit a gluon, and the quark current $J_{ij\pm}$ can destroy or produce a quark-antiquark pair, or destroy and create a quark or antiquark. 

The helicity operator basis has simple behavior under discrete symmetries. For example, 
\begin{equation} \label{eq:Cfield}
\C\, \cB^a_{i\pm}\, T^a_{\alpha\bbeta}\,\C = - \cB^a_{i \pm} T^a_{\beta\balpha}
\,, \quad
\C\, J^{\balpha\beta}_{ij\pm}\,\C = -J^{\bbeta\alpha}_{ji\mp}
\,,\end{equation}
Charge conjugation and parity invariance reduce the number of independent Wilson coefficients. 

Since the polarizations of gluons can be treated in $d$ rather than 4 dimensions, it is natural to ask whether our helicity operator basis is complete. Operators with $\eps$-dimensional polarizations do arise in the matching for states with physical polarizations. They are also not generated by the renormalization group evolution: The only communication between collinear sectors is through soft radiation, which does not carry spin and therefore cannot change helicity. 

\section{Conclusions and outlook}

We have described a helicity operator basis, that makes it straightforward to write down the complete basis for a hard scattering process. It also facilitates the matching from fixed-order calculations onto SCET, since the matching coefficients are directly given in terms of the color-ordered helicity amplitudes. We demonstrated its ease by obtaining the Wilson coefficients for $pp\to H + 0,1,2$ jets, $pp\to W/Z/\gamma + 0,1,2$ jets, and $pp\to 2,3$ jets at (next-to-)leading order~\cite{Moult:2015aoa}.

The spin of the operators does not play a crucial role at leading power, as the helicities are simply summed over in \eq{sigma}. This is not true for color, since soft gluons can exchange color. However, at subleading power also the spin structure is essential, since soft gluons can then also transfer spin. Our helicity approach was key in constructing a basis of subleading operators~\cite{Kolodrubetz:2016uim}.

Spin information also needs to be kept track of when matching between different SCET theories. For example, to describe two nearby hard jets one matches through an intermediate SCET where the two nearby jets are not separately resolved~\cite{Bauer:2011uc}. To keep track of this spin information in the matching, helicity fields proved particularly useful~\cite{Pietrulewicz:2016nwo}.

\section*{Acknowledgments}

This work was supported in part by the Office of Nuclear Physics of the U.S.
Department of Energy under Contract No. DE-SC0011090,
the DFG Emmy-Noether Grant No. TA 867/1-1, 
the Marie Curie International Incoming Fellowship PIIF-GA-2012-328913,
the Simons Foundation Investigator Grant No. 327942,
NSERC of Canada,
and the D-ITP consortium, a program of the Netherlands Organization for Scientific Research (NWO) that is funded by the Dutch Ministry of Education, Culture and Science (OCW).

\section*{References}

\end{document}